\documentclass[preprint,eqsecnum,aps,nofootinbib]{revtex4}
\usepackage[dvips]{graphicx}
\usepackage{amssymb}
\usepackage{amsmath}
\usepackage{amsmath, amsthm, amssymb, mathrsfs}
\usepackage{pstricks}

\newcommand{\ket}[1]{\lvert #1 \rangle}
\newcommand{\bra}[1]{\langle #1 \lvert}
\begin{document}

\title{Attack of Many Eavesdroppers via Optimal Strategy in Quantum Cryptography} 

\author{Eylee Jung, Mi-Ra Hwang, DaeKil Park}
\affiliation{Department of Physics, Kyungnam University, Masan,
631-701, Korea}
\author{Hungsoo Kim}
\affiliation{Department of Applied Mathematics, Pukyong National University, Pusan, 606-737,
Korea}
\author{Eui-Soon Yim}
\affiliation{Department of Computer Science, Semyung University, Chechon, 390-711, Korea}
\author{Jin-Woo Son}
\affiliation{Department of Mathematics, Kyungnam University, Masan,
631-701, Korea  }

%\date{\today}
\begin{abstract}
We examine a situation that $n$ eavesdroppers attack the Bennett-Brassard cryptographic protocol
via their own optimal and symmetric strategies.
Information gain and mutual information with sender for each eavesdropper are explicitly 
derived. The receiver's error rate for the case of arbitrary $n$ eavesdroppers 
can be derived using a recursive relation.
Although first eavesdropper can get mutual information without disturbance arising due to 
other eavesdroppers, subsequent eavesdropping generally increases the receiver's error rate.
Other eavesdroppers cannot gain information on the input signal sufficiently. As a result, 
the information each eavesdropper gains becomes less than optimal one.
\end{abstract}

%\pacs{04.20.Ha, 04.20.Jb, 11.27.+d}

\maketitle

%---------------------------------------------------------------------------
\section{Introduction}
Quantum cryptography is one of the major applications of quantum information 
theories\cite{nielsen00,Peres} While other applications such as quantum teleportation and quantum
computer require tens or even thousands of qubits, the quantum cryptography scenario such
as BB84 protocol\cite{bb84} can be implemented, at least theoretically, using only single 
qubit technology. This is main reason why the quantum cryptography based on BB84 or 
Ekert91\cite{ek91} is now at the stage of the industrial era\cite{all07}.

According to the usual BB84 protocol the sender (Alice)  sends a single qubit to 
the receiver (Bob) by choosing randomly one of the conjugate bases $\{ \ket{x}, \ket{y} \}$ and 
$\{ \ket{u}, \ket{v} \}$, where 
\begin{equation}
\label{conjugate1}
\ket{u} = \frac{1}{\sqrt{2}} \left( \ket{x} + \ket{y} \right)    \hspace{1.0cm}
\ket{v} = \frac{1}{\sqrt{2}} \left( \ket{x} - \ket{y} \right). 
\end{equation}
Then Bob performs a quantum mechanical measurements in these bases. After measurements, Alice and
Bob communicate with each other via classical public channel and establish a secret quantum key
by using only those cases in which the bases of Alice and Bob coincide.

How much information an eavesdropper (Eve) can gain when Alice and Bob perform the usual BB84
scheme? The answer of this question is important to check the security of the quantum 
cryptography. In this reason many authors examined the various strategies with one- and
two-dimensional probes\cite{hutt94,lutk96,gisin97,fuchs97}. Among them Ref.\cite{fuchs97} derived
the optimal (or maximal) mutual information between Alice and Eve as a function of the 
disturbance $D$ in the BB84 protocol. The final result can be summarized as follows:
\begin{equation}
\label{optimal-1}
{\cal I}_{xy} = \frac{1}{2} \phi \left[2 \sqrt{D_{uv} (1 - D_{uv})} \right]  \hspace{1.0cm}
{\cal I}_{uv} = \frac{1}{2} \phi \left[2 \sqrt{D_{xy} (1 - D_{xy})} \right],
\end{equation}
where ${\cal I}_{xy}$ (or ${\cal I}_{uv}$) is the optimal mutual 
information when Alice sends a signal
to Bob via $x-y$ (or $u-v$) basis, and $\phi(z) = (1+z) \log_2 (1+z) + (1-z) \log_2 (1-z)$.
The constants $D_{xy}$ and $D_{uv}$ denote the disturbances in these bases. The most different
point of the quantum cryptography from the classical one is the fact that Eve cannot get
information from the trusted parties without arising the disturbance. This implies that the
quantum scheme is more secure than the classical cryptography.

\begin{figure}[ht!]
\begin{center}
\includegraphics[height=8.0cm]{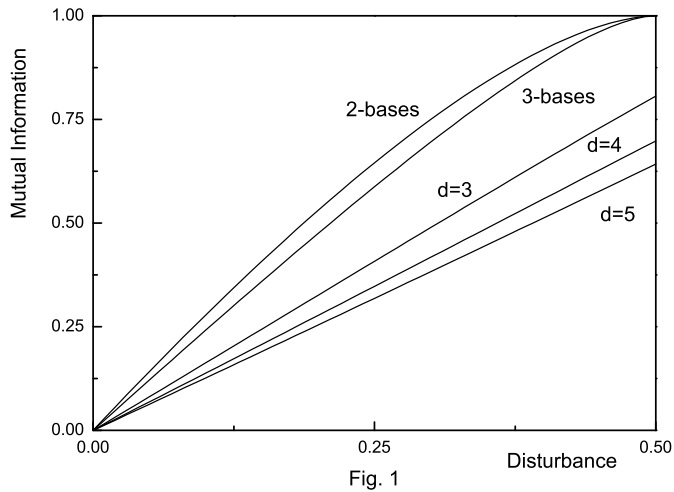}
\caption[fig1]{Plot of $D$-dependence of the optimal mutual information when Alice and Bob
use the various different protocols.}
\end{center}
\end{figure}

Recently, many different cryptographic protocols have been studied from the purely theoretical
ground (at least at current stage) even if most quantum cryptography has been demonstrated by
making use of either one of BB84 or Ekert91 protocols. One of the motivations for searching 
other protocols is to strength the security against eavesdropping. The simple extended 
protocol is a six-state protocol\cite{bruss98,bech99}. In this protocol Alice sends a signal to 
Bob after choosing randomly one of three conjugate bases $\{ \ket{x}, \ket{y} \}$, 
$\{ \ket{u}, \ket{v} \}$ and $\{ \ket{w}, \ket{z} \}$, where
\begin{equation}
\label{conjugate2}
\ket{w} = \frac{1}{\sqrt{2}} \left( \ket{x} + i \ket{y} \right)   \hspace{1.0cm}
\ket{z} = \frac{1}{\sqrt{2}} \left( \ket{x} - i \ket{y} \right).
\end{equation}
The basis $\{ \ket{w}, \ket{z} \}$ corresponds to the circular polarization if Alice and Bob 
use a photon polarization as a qubit. The optimal mutual information between Alice and Eve 
is plotted in Fig. 1, which implies that the six-state protocol is more secure than usual
four-state BB84 against eavesdropping. Another extended 
protocol\cite{bech00,bruss02,boure01,cerf01} is that Alice and Bob use qutrit ($d=3$) or 
more generally qudit ($d=4, 5, \cdots$) instead of a qubit. The optimal mutual information in
this protocol is also plotted in Fig. 1 when $d=3$, $4$, and $5$. Fig. 1 indicates that the 
protocol with $d$-level system is more secure against eavesdropping with increasing $d$.
Furthermore, the quantum cryptography with continuum\cite{piran08} and noisy states\cite{shad08}
are under investigation. However, all of these other protocols seem to be far from embodiment 
in a few years from the aspect of experimental science.

In this paper we would like to explore the situation where many eavesdroppers 
(Eve1, Eve2, $\cdots$) attack the BB84 protocol optimally. We assume that all of the eavesdroppers
think they are unique eavesdropper. Our computation is based on the quantum circuit expression
of the optimal eavesdropping strategy\cite{griffi97}. This paper is organized as follows. In 
section II we review Ref.\cite{griffi97} briefly. In this section we develop a computational 
technique, which is useful when many eavesdroppers try to attack optimally.
In section III we examine the situation where 
Eve1 and Eve 2 attack the usual BB84 protocol. Information gain $G^{(i)}$ and mutual information
$I^{(i)}$ are explicitly computed, where $i=1$ or $2$ corresponds 
to Eve1 and Eve2 respectively. When
Eve1 and Eve2 attack via symmetric optimal strategy, we compute the Bob's error rate or 
disturbance $D_{B,2}$ explicitly, where the subscript ``2'' denotes the two eavesdroppers. 
It turns out that both optimal strategies fail. Although Eve1 can gain information on 
Alice's signal as much as possible, Eve2 increases the disturbance or Bob's error rate. For 
Eve2 she cannot gain information sufficiently due to Eve1's disturbance. 
In section IV we examine the situation where three eavesdroppers attack the BB84 protocol.
The mutual information for each eavesdropper is analytically derived. Furthermore, Bob's 
error rate $D_{B,3}$ is also explicitly derived on condition that all eavesdroppers use the 
symmetric strategies.
In section V we have generalized the results of the previous sections. When $n$ eavesdroppers 
attack, the mutual information for each eavesdropper is analytically derived. Also the 
recursive relation of the Bob's error rate is derived. It turns out that all optimal 
strategies eavesdroppers choose eventually fail except very rare cases. Finally
a brief concluding remark is given.

\section{One Eavesdropper}

%%%%%%%%%%%%%%%%%%%%%%%%%%%%%%%%%%%%%%%%%%%%%%%%%%%%%%%%%%%%%%%%%%%%%%%%%%%%%%%%%%
\begin{figure}
\begin{center}
\begin{pspicture}(0, 0)(11, 2.5)
%%%%%%%%%%  GATES

%%%%%%%%%%  Horizontal LINES
   \psline[linewidth=0.7pt]( 1.00,  2.00)( 4.00,  2.00)%
   \psline[linewidth=0.7pt]( 1.00,  1.00)( 4.00,  1.00)%
   \psline[linewidth=0.7pt]( 1.00,  0.50)( 4.00,  0.50)%
   
   \psline[linewidth=0.7pt]( 7.00,  2.00)( 10.00,  2.00)%
   \psline[linewidth=0.7pt]( 7.00,  1.00)( 10.00,  1.00)%
   \psline[linewidth=0.7pt]( 7.00,  0.50)( 10.00,  0.50)%
   
%%%%%%%%%%  Vertical LINES
   \psline[linewidth=0.7pt]( 2.00,  2.00)( 2.00,  1.00)%
   \psline[linewidth=0.7pt]( 3.00,  2.00)( 3.00,  0.50)%
   
   \psline[linewidth=0.7pt]( 8.00,  2.00)( 8.00,  1.00)%
   \psline[linewidth=0.7pt]( 9.00,  2.00)( 9.00,  0.50)%
   
%%%%%%%%% DOTS
 \psdot[dotstyle=*, dotscale=2]( 2.00,   2.00)%
   \psdot[dotstyle=*, dotscale=2]( 3.00,   0.50)%
   \psdot[dotstyle=oplus, dotscale=2]( 2.00,   1.00)%
   \psdot[dotstyle=oplus, dotscale=2]( 3.00,   2.00)%
   
   \psdot[dotstyle=oplus, dotscale=2]( 8.00,   2.00)%
   \psdot[dotstyle=oplus, dotscale=2]( 9.00,   0.50)%
   \psdot[dotstyle=*, dotscale=2]( 8.00,   1.00)%
   \psdot[dotstyle=*, dotscale=2]( 9.00,   2.00)%
   
%%%%%%%%%%  TEXTS
   \rput[r]( 0.80,  2.00){Alice}%
   \rput[l]( 4.20,  2.00){Bob}%
   \rput[r]( 0.90,  1.00){$\lvert e_{0} \rangle$}
   \rput[r]( 0.90,  0.50){$\lvert f_{0} \rangle$}
   \rput[c]( 2.50,  0.00){(a)}%
   
   \rput[r]( 6.80,  2.00){Alice}%
   \rput[l](10.20,  2.00){Bob}%
   \rput[r]( 6.90,  1.00){$\lvert e_{0} \rangle$}
   \rput[r]( 6.90,  0.50){$\lvert f_{0} \rangle$}
   \rput[c]( 8.50,  0.00){(b)}%

\end{pspicture}
\end{center}
\caption[fig2]{Quantum circuit expression for the optimal eavesdropping strategy. 
Fig. 2(a) and (b) 
represent the optimal strategy when Alice sends a signal using $x-y$ or $u-v$ basis respectively.
The bottom two lines belong to Eve and the top line to Alice. Time advances from left to right.}
\end{figure}
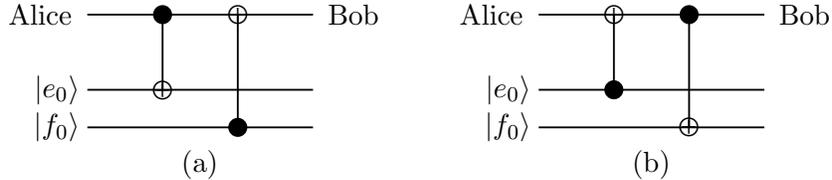
%%%%%%%%%%%%%%%%%%%%%%%%%%%%%%%%%%%%%%%%%%%%%%%%%%%%%%%%%%%%%%%%%%%%%%%%%

The quantum circuits for the optimal eavesdropping in $x-y$ and $u-v$ bases are given in Fig. 2.
The top line belongs to Alice and Bob, and the bottom two lines to Eve. In order to perform
the optimal eavesdropping strategy Eve prepares the initial states as following:
\begin{eqnarray}
\label{1optimal-1}
& &\ket{e_0} = \sqrt{1 - \Delta_{uv}} \ket{x} + \sqrt{\Delta_{uv}} \ket{y} = 
\sqrt{1 - D_{uv}} \ket{u} + \sqrt{D_{uv}} \ket{v}   
                                                           \\    \nonumber
& &\ket{f_0} = \sqrt{1 - D_{xy}} \ket{x} + \sqrt{D_{xy}} \ket{y} = 
\sqrt{1 - \Delta_{xy}} \ket{u} + \sqrt{\Delta_{xy}} \ket{v}
\end{eqnarray}
where $\Delta$ and $D$ are related, when they have same subscripts, through the formula
\begin{equation}
\label{formula-1}
\Delta = \frac{1}{2} - \sqrt{D (1 - D)}     \hspace{1.0cm}
D = \frac{1}{2} - \sqrt{\Delta (1 - \Delta)}.
\end{equation}

If Alice sends a signal using $x-y$ basis, Fig. 2(a) shows that the entangled states 
between Alice and Eve becomes  
\begin{eqnarray}
\label{1optimal-2}
& &\ket{x} \rightarrow \ket{X} = \alpha_0 \ket{xxx} + \alpha_1 \ket{yxy} + \alpha_2 \ket{xyx}
+ \alpha_3 \ket{yyy}             \\   \nonumber
& & \ket{y} \rightarrow \ket{Y} = \alpha_0 \ket{yyx} + \alpha_1 \ket{xyy} + \alpha_2 \ket{yxx} +
\alpha_3 \ket{xxy}
\end{eqnarray}
where
\begin{eqnarray}
\label{1optimal-3}
& &\alpha_0 = \sqrt{1 - \Delta_{uv}} \sqrt{1 - D_{xy}} \hspace{1.0cm}
\alpha_1 = \sqrt{1 - \Delta_{uv}} \sqrt{D_{xy}}   \\    \nonumber
& &\alpha_2 = \sqrt{\Delta_{uv}} \sqrt{1 - D_{xy}}   \hspace{1.0cm}
\alpha_3 = \sqrt{\Delta_{uv}} \sqrt{\Delta_{xy}}.
\end{eqnarray}

For later use it is necessary to express Eq.(\ref{1optimal-2}) more compactly. This can be 
achieved by 
\begin{equation}
\label{compact-1}
\ket{X} = \sum_{i=0}^3 \alpha_i \ket{i}_2 \ket{i}_4  \hspace{1.0cm}
\ket{Y} = \sum_{i=0}^3 \alpha_i \ket{i+1}_2 \ket{i+2}_4
\end{equation}
where $\ket{j}_2$ and $\ket{j}_4$ means $\ket{j \hspace{.2cm} \mbox{modulo} \hspace{.2cm}2}$ and 
$\ket{j \hspace{.2cm} \mbox{modulo} \hspace{.2cm} 4}$. Thus $\ket{j}_2$ and $\ket{j}_4$ represents
the one- and two-qubit states respectively with ordering $x$ and $y$ for $\ket{j}_2$ and 
$xx$, $xy$, $yx$ and $yy$ for $\ket{j}_4$. This compact notation will be usefully used in the 
following sections when many eavesdroppers attack.

When Alice sends a signal using $u-v$ basis, the usual control-NOT gate changes
\begin{equation}
\label{1uv-1}
\ket{uu} \rightarrow \ket{uu}  \hspace{.5cm}
\ket{uv} \rightarrow \ket{vv}  \hspace{.5cm}
\ket{vu} \rightarrow \ket{vu}  \hspace{.5cm}
\ket{vv} \rightarrow \ket{uv}.
\end{equation}
Thus the control-NOT gate in $x-y$ basis can be easily understood in $u-v$ basis by exchanging
the control gate with target gate. This is a reason why Fig. 2(b) used in $u-v$ basis is 
different from Fig. 2(a). 

Now, we want to show that the entangled states (\ref{1optimal-2}) with suitable POVM
measurement enables Eve to get information optimally. The complete set of the positive
operators, which is used for POVM, can be derived generally as projective operators onto the 
eigenvectors of $\Gamma_{xy} = \rho_x - \rho_y$, where\cite{fuchs96}
\begin{equation}
\label{1POVM-1}
\rho_x = \mbox{Tr}_{Alice} \ket{X} \bra{X}   \hspace{1.0cm}
\rho_y = \mbox{Tr}_{Alice} \ket{Y} \bra{Y}.
\end{equation}
For our case the complete set of the positive operators is $\{ E_0, E_1, E_2, E_3 \}$ with
$E_0 = \ket{xx}\bra{xx}$, $E_1 = \ket{xy}\bra{xy}$, $E_2 = \ket{yx}\bra{yx}$, and 
$E_3 = \ket{yy}\bra{yy}$. Then it is easy to compute 
$P_{\lambda i} = \bra{I} \openone \otimes E_{\lambda} \ket{I}$ with $I = X$ or $Y$ and 
$i=x$ or $y$, which is 
the probability that Eve detects outcome $\lambda$ when Alice sends a signal $i$:
\begin{eqnarray}
\label{1optimal-4}
& &P_{0x} = \alpha_0^2 \hspace{.5cm} P_{1x} = \alpha_1^2 \hspace{.5cm}
P_{2x} = \alpha_2^2 \hspace{.5cm} P_{3x} = \alpha_3^2              \\   \nonumber
& &P_{0y} = \alpha_2^2 \hspace{.5cm} P_{1y} = \alpha_3^2 \hspace{.5cm}
P_{2y} = \alpha_0^2 \hspace{.5cm} P_{3y} = \alpha_1^2. 
\end{eqnarray}
Using Eq.(\ref{1optimal-4}), one can
compute $q_{\lambda} = (1/2) (P_{\lambda x} + P_{\lambda y})$ and 
$Q_{i \lambda} = (1/2) P_{\lambda i} / q_{\lambda}$:
\begin{equation}
\label{1optimal-5}
q_0 = \frac{1}{2} (1 - D_{xy})  \hspace{.5cm} q_1 = \frac{1}{2} D_{xy}  \hspace{.5cm}
q_2 = \frac{1}{2} (1 - D_{xy})  \hspace{.5cm} q_3 = \frac{1}{2} D_{xy}
\end{equation}
and 
\begin{eqnarray}
\label{1optimal-6}
& &Q_{x0} = 1 - \Delta_{uv} \hspace{.5cm} Q_{x1} = 1 - \Delta_{uv} \hspace{.5cm}
Q_{x2} = \Delta_{uv} \hspace{.5cm} Q_{x3} = \Delta_{uv}
                                                      \\    \nonumber
& &Q_{y0} = \Delta_{uv} \hspace{.5cm} Q_{y1} = \Delta_{uv} \hspace{.5cm}
Q_{y2} = 1 - \Delta_{uv} \hspace{.5cm} Q_{y3} = 1 - \Delta_{uv}.
\end{eqnarray}
The quantity $q_{\lambda}$ is a probability that Eve has outcome $\lambda$ when Alice uses 
$x-y$ basis. The quantity $Q_{i \lambda}$ is posterior probability on the Eve's guess after
she has a outcome $\lambda$. Then the information gain is defined as 
$G_{\lambda} = |Q_{x \lambda} - Q_{y \lambda}|$, which, for our case, is 
$\lambda$-independent:
\begin{equation}
\label{1optimal-7}
G_{\lambda} = 1 - 2 \Delta_{uv} = 2 \sqrt{D_{uv} (1 - D_{uv})}.
\end{equation}
Thus, the mutual information ${\cal I}^{AE}$ between Alice and Eve reduces to
\begin{equation}
\label{1optimal-8}
{\cal I}^{AE} \equiv \frac{1}{2} \sum_{\lambda} q_{\lambda} \phi (G_{\lambda}) = 
\frac{1}{2} \phi \left[2 D_{uv} (1 - D_{uv}) \right]
\end{equation}
where $\phi(z) = (1+z) \log_{2} (1+z) + (1-z) \log_{2} (1-z)$.

Now, let us derive the Bob's error rate, usually called disturbance when Alice sends a 
signal using the $x-y$ basis. First, we consider the following quantities:
\begin{equation}
\label{1optimal-9}
d_{\lambda u} \equiv 1 - \frac{\bra{U} B_u \otimes E_{\lambda} \ket{U}}
                              {\bra{U} \openone \otimes E_{\lambda} \ket{U}}   \hspace{1.0cm}
d_{\lambda v} \equiv 1 - \frac{\bra{V} B_v \otimes E_{\lambda} \ket{V}}
                              {\bra{V} \openone \otimes E_{\lambda} \ket{V}}
\end{equation}
where $\ket{U} = (1 / \sqrt{2}) (\ket{X} + \ket{Y})$, 
$\ket{V} = (1 / \sqrt{2}) (\ket{X} - \ket{Y})$, $B_u = \ket{u}\bra{u}$, and $B_v = \ket{v}\bra{v}$.
These are probabilities Bob gets a wrong result {\it conditioned upon} Alice sending 
$\ket{u}$ or $\ket{v}$, and Eve measuring $\lambda$. Computation of $d_{\lambda u}$ and 
$d_{\lambda v}$ is straightforward. The result is that $d_{\lambda u}$ is identical to 
$d_{\lambda v}$ and they are also $\lambda$-independent as follows:
\begin{equation}
\label{1optimal-10}
d_{\lambda u} = d_{\lambda v} \equiv d_{\lambda} = D_{uv}. \hspace{.5cm} (\lambda = 0,1,2,3)
\end{equation}
Then the Bob's error rate $D_B$ is given by 
\begin{equation}
\label{1optimal-11}
D_B \equiv \sum_{\lambda} q_{\lambda} d_{\lambda} = D_{uv}.
\end{equation}
Thus Eq.(\ref{1optimal-8}) can be re-written as 
\begin{equation}
\label{1optimal-12}
{\cal I}^{AE} = \frac{1}{2} \phi \left[2 D_B (1 - D_B)\right]
\end{equation}
which is the optimal mutual information derived in Ref.\cite{fuchs97} when Alice sends a signal
using $x-y$ basis. If Alice uses $u-v$ basis, we should repeat the previous calculation using 
Fig. 2(b). The final result is identical with Eq.(\ref{1optimal-12}) except $D_B = D_{xy}$. Thus,
the strategies expressed by Fig. 2 give an optimal information to Eve regardless of 
the basis Alice is using.

\section{Two Eavesdroppers}

%%%%%%%%%%%%%%%%%%%%%%%%%%%%%%%%%%%%%%%%%%%%%%%%%%%%%%%%
\begin{figure}
\begin{center}
\begin{pspicture}(0, 0)(6, 4)
%%%%%%%%%%  GATES

%%%%%%%%%%  Horizontal LINES
   \psline[linewidth=0.7pt]( 1.00,  3.50)( 5.00,  3.50)%
   
   \psline[linewidth=0.7pt]( 1.00,  2.50)( 5.00,  2.50)%
   \psline[linewidth=0.7pt]( 1.00,  2.00)( 5.00,  2.00)%
   
   \psline[linewidth=0.7pt]( 1.00,  1.00)( 5.00,  1.00)%
   \psline[linewidth=0.7pt]( 1.00,  0.50)( 5.00,  0.50)%
%%%%%%%%%%  Vertical LINES
   \psline[linewidth=0.7pt]( 2.00,  3.50)( 2.00,  2.50)%
   \psline[linewidth=0.7pt]( 2.50,  3.50)( 2.50,  2.00)%
   
   \psline[linewidth=0.7pt]( 3.50,  3.50)( 3.50,  1.00)%
   \psline[linewidth=0.7pt]( 4.00,  3.50)( 4.00,  0.50)%
   
   \psline[linewidth=0.5pt, linestyle=dashed]( 4.50,  3.80)( 4.50,  0.20)%
%%%%%%%%% DOTS
   \psdot[dotstyle=*, dotscale=2]( 2.00,   3.50)%
   \psdot[dotstyle=*, dotscale=2]( 2.50,   2.00)%
   \psdot[dotstyle=oplus, dotscale=2]( 2.00,   2.50)%
   \psdot[dotstyle=oplus, dotscale=2]( 2.50,   3.50)%

   \psdot[dotstyle=*, dotscale=2]( 3.50,   3.50)%
   \psdot[dotstyle=*, dotscale=2]( 4.00,   0.50)%
   \psdot[dotstyle=oplus, dotscale=2]( 3.50,   1.00)%
   \psdot[dotstyle=oplus, dotscale=2]( 4.00,   3.50)%
%%%%%%%%%%  TEXTS
   \rput[r]( 0.80,  3.50){Alice}%
   \rput[l]( 5.20,  3.50){Bob}%
   \rput[r]( 0.90,  2.50){$\scriptstyle{\lvert e_{0}^{(1)} \rangle}$}
   \rput[r]( 0.90,  2.00){$\scriptstyle{\lvert f_{0}^{(1)} \rangle}$}

   \rput[r]( 0.90,  1.00){$\scriptstyle{\lvert e_{0}^{(2)} \rangle}$}
   \rput[r]( 0.90,  0.50){$\scriptstyle{\lvert f_{0}^{(2)}\rangle}$}
%%%%%%%%%%%%%%%%%%%%%%%%%%%%%%%%%%%%%

\end{pspicture}
\end{center}
\caption[fig3]{Quantum circuit expression for the situation where two eavesdroppers Eve1 and Eve2
attack the usual BB84 protocol when Alice sends a signal to Bob using $x-y$ basis. The top
line belongs to Alice, next two lines to Eve1, and bottom two lines to Eve2. If Alice uses a
$u-v$ basis, this figure should be modified by exchanging the control gates with target 
gates in all control-NOT gates. Time advances from left to right.}
\end{figure}
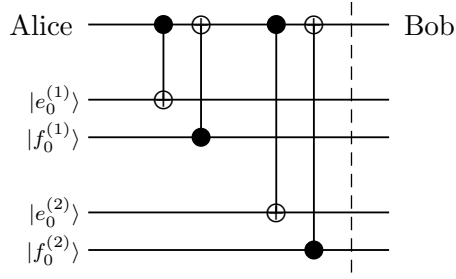
%%%%%%%%%%%%%%%%%%%%%%%%%%%%%%%%%%%%%%%%%%%%%%%%%%%%%%%%%%%

Now we consider a situation that two eavesdroppers, Eve1 and Eve2, attack the usual BB84 protocol.
We assume that Eve1 and Eve2 do not know each other and they use their own optimal strategies.
Thus corresponding quantum circuit should be Fig. 3 when Alice sends a signal using $x-y$
basis. From now on we will use the superscript $(i)$ to distinguish the quantities (or states)
which belong to Eve1 and Eve2.

Using the compact notation used in Eq.(\ref{compact-1}), one can derive the entangled states
at the stage represented as a dotted line in Fig. 3:
\begin{eqnarray}
\label{compact-2}
& &\ket{x} \rightarrow \ket{X} = \sum_{i,j=0}^3 \alpha_i^{(1)} \alpha_j^{(2)} 
\ket{i+j}_2 \ket{i}_4 \ket{2i+j}_4                  \\   \nonumber
& &\ket{y} \rightarrow \ket{Y} = \sum_{i,j=0}^3 \alpha_i^{(1)} \alpha_j^{(2)} 
\ket{i+j+1}_2 \ket{i+2}_4 \ket{2i+j+2}_4.
\end{eqnarray}

In order to derive the POVM elements for Eve1 we construct the operator
\begin{equation}
\label{2POVM-1}
\Gamma_{xy}^{(1)} = \rho_x^{(1)} - \rho_y^{(1)}
\end{equation}
where
\begin{equation}
\label{2POVM-2}
\rho_x^{(1)} = \mbox{Tr}_{A,E2} \ket{X}\bra{X}   \hspace{1.0cm}
\rho_y^{(1)} = \mbox{Tr}_{A,E2} \ket{Y}\bra{Y}.
\end{equation}
In Eq.(\ref{2POVM-2}) $\mbox{Tr}_{A,E2}$ means a partial trace over Alice and Eve2's qubits.  
Then it is easy to compute the eigenvectors of $\Gamma_{xy}^{(1)}$, which gives the complete
set of the positive operators $\{E_0^{(1)}, E_1^{(1)}, E_2^{(1)}, E_3^{(1)}\}$ to Eve1, where
\begin{equation}
\label{2POVM-3}
E_0^{(1)} = \ket{xx}_{2,3}\bra{xx}    \hspace{.5cm}
E_1^{(1)} = \ket{xy}_{2,3}\bra{xy}    \hspace{.5cm}
E_2^{(1)} = \ket{yx}_{2,3}\bra{yx}    \hspace{.5cm}
E_3^{(1)} = \ket{yy}_{2,3}\bra{yy}.
\end{equation}   
The subscript $2,3$ means qubits of second and third lines in Fig. 3. By same way one can 
construct the complete set of the positive operators for Eve2, which is 
\begin{equation}
\label{2POVM-4}
E_0^{(2)} = \ket{xx}_{4,5}\bra{xx}    \hspace{.5cm}
E_1^{(2)} = \ket{xy}_{4,5}\bra{xy}    \hspace{.5cm}
E_2^{(2)} = \ket{yx}_{4,5}\bra{yx}    \hspace{.5cm}
E_3^{(2)} = \ket{yy}_{4,5}\bra{yy}.  
\end{equation}

Then the remaining calculation for the mutual informations ${\cal I}^{A E_1}$ between Alice and 
Eve1, and ${\cal I}^{A E_2}$ between Alice and Eve2 is straightforward. The information gains
$G_{\lambda}^{(1)}$ for Eve1 and $G_{\lambda}^{(2)}$ for Eve2 turn out to be 
$\lambda$-independent as following:
\begin{equation}
\label{2gain-1}
G_{\lambda}^{(1)} = 1 - 2 \Delta_{uv}^{(1)}   \hspace{1.0cm}
G_{\lambda}^{(2)} = (1 - 2 \Delta_{uv}^{(2)}) (1 - 2 D_{xy}^{(1)}).   
\hspace{1.0cm}  (\lambda = 0, 1, 2, 3)
\end{equation}
Therefore from a comparison of Eq.(\ref{2gain-1}) with Eq.(\ref{1optimal-7}) 
Eve1 seems to be able to 
get information as much as the case of unique eavesdropper. 
This is due to the fact that Eve1 attacks the
BB84 protocol earlier than Eve2 and therefore, gathers information without perturbation arising
due to Eve2. However, this does not mean that Eve1's optimal strategy is succeeded. As shown in
Fig. 1 optimality of the eavesdropping does not uniquely depend on the quantity of information
that eavesdropper can gain. 
In order to get success in the eavesdropping, eavesdropper should decrease
the disturbance as much as possible. These two factors, increase of information gain and decrease
of disturbance, determine the success or failure of the optimal strategy. As will be shown
shortly, Eve1's optimal strategy fails because Eve2 increases Bob's error rate.
For Eve2 the information gain involves an interesting factor $1 - D_{xy}^{(1)}$. Thus Eve2's 
information gain depends on the Eve1's choice of $D_{xy}^{(1)}$. This is manifestly due to the 
fact that Eve2 performs her optimal strategy after Eve1. If Eve1 chooses $D_{xy}^{(1)} = 0$,
Eve2 can get information as much as Eve1 if $\Delta_{uv}^{(1)} = \Delta_{uv}^{(2)}$. 
This indicates that Eve2 can increase her information gain if Eve1 does not disturb the signal
Alice sent to Bob. The mutual information ${\cal I}^{A E_1}$ and ${\cal I}^{A E_2}$ reduce to 
\begin{equation}
\label{2mutual-1}
{\cal I}^{A E_1} = \frac{1}{2} \phi \left( G_{\lambda}^{(1)} \right)  \hspace{1.0cm}
{\cal I}^{A E_2} = \frac{1}{2} \phi \left( G_{\lambda}^{(2)} \right).
\end{equation}

Now, let us turn into the Bob's error rate. Unlike the unique eavesdropper case discussed in
the previous section the situation is very complicated. In this case it could happen that 
Eve1's disturbance and Eve2's successive disturbance does not generate an error to Bob.
Thus equation corresponding to Eq.(\ref{1optimal-9}) in previous section should have one more 
index, i.e. $d_{\lambda u} \rightarrow d_{\lambda \lambda' u}$ and 
$d_{\lambda v} \rightarrow d_{\lambda \lambda' v}$. Since, furthermore, both optimal strategies
Eve1 and Eve2 have chosen do not get success, we expect to have 
$d_{\lambda \lambda' u} \neq d_{\lambda \lambda' v}$.
Thus we should compute the Bob's error rate separately when Alice sends $\ket{u}$ and 
$\ket{v}$. Since computation in this way needs long and tedious calculation, we will try
to make the situation more simple.

To make the situation more simple we assume that both eavesdropping strategies are symmetric,
i.e. $D_{xy}^{(1)} = D_{uv}^{(1)}$ and $D_{xy}^{(2)} = D_{uv}^{(2)}$. In this case we can 
compute the Bob's error rate directly from the entangled states Eq. (\ref{compact-2}), which is 
\begin{equation}
\label{2Bob-1}
D_B = D^{(1)} (1 - D^{(2)}) + D^{(2)} (1 - D^{(1)}).
\end{equation}
In Eq.(\ref{2Bob-1}) we omit the subscript because it is useless in the symmetric strategies. 
If $D^{(2)} = 0$, $D_B$ becomes $D^{(1)}$ which is Bob's error rate if Eve1 is an unique 
eavesdropper. If $D^{(1)} = 0$, $D_B$ becomes $D^{(2)}$ which is also Bob's error rate if
Eve2 is an unique eavesdropper. The general Bob's error rate becomes nice combination of 
$D^{(1)}$ and $D^{(2)}$.

\begin{figure}[ht!]
\begin{center}
\includegraphics[height=6.5cm]{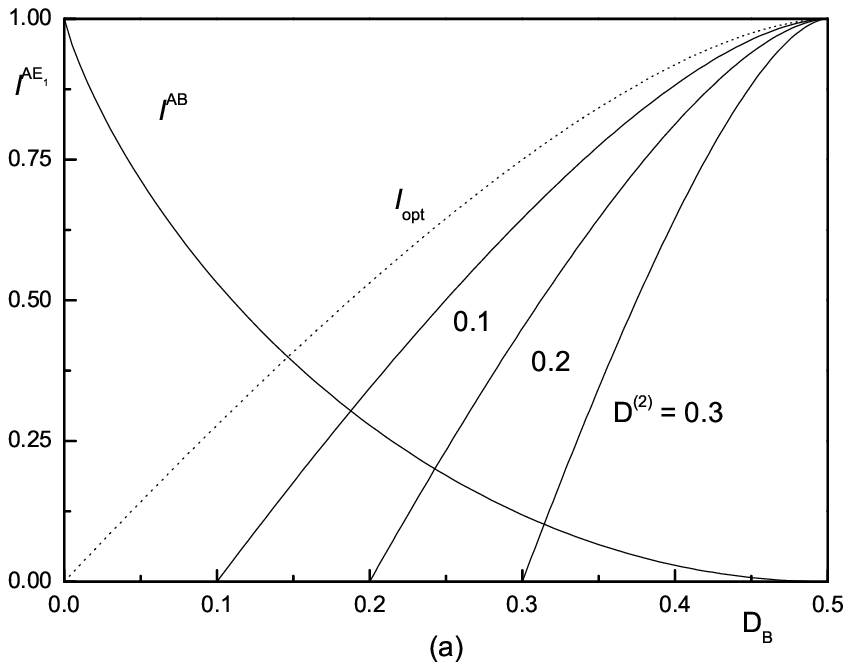}
\includegraphics[height=6.5cm]{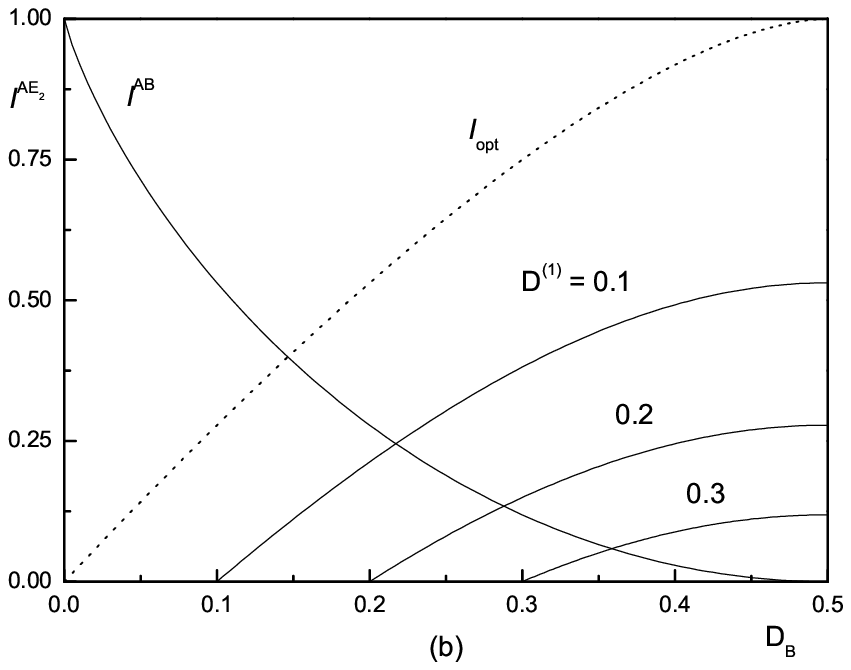}
\caption[fig4]{Plot of $D_B$-dependence of ${\cal I}^{A E_1}$ (Fig. 4(a)) and 
${\cal I}^{A E_2}$ (Fig. 4(b)). The dotted line is a $D_B$-dependence of the optimal strategy
derived in Eq.(\ref{optimal-1}). The monotonically decreasing line corresponds to 
${\cal I}^{AB}$, mutual information between Alice and Bob. Fig. 4(a) implies that the mutual 
information of Eve1 is less than the optimal one except $D^{(2)} = 0$. This is due to the fact
that Eve2's eavesdropping process generally increases the Bob's error rate. Fig. 4(b) implies 
that the mutual information of Eve2 is also less than the optimal one except $D^{(1)} = 0$. This
is due to the fact that Eve1's eavesdropping process generally decreases the information gain
for Eve1.}
\end{center}
\end{figure}

Fig. 4 is $D_B$-dependence of ${\cal I}^{A E_1}$ (Fig. 4(a)) and ${\cal I}^{A E_2}$ (Fig. 4(b)).
We plot ${\cal I}^{A E_1}$ in Fig. 4(a) when $D^{(2)} = 0.1$, $0.2$ and $0.3$ respectively.
For comparison we plot the optimal information ${\cal I}_{opt}$ (see Eq.(\ref{optimal-1})) and 
mutual information ${\cal I}^{AB}$ between Alice and Bob defined
\begin{equation}
\label{2Bob-2}
{\cal I}^{AB} = 1 + D_B \log_2 D_B + (1 - D_B) \log_2 (1 - D_B)
\end{equation}
together. As Fig. 4(a) indicates, Eve1's mutual information with Alice are in general smaller 
than ${\cal I}_{opt}$ when $D^{(2)} \neq 0$. If $D^{(2)}$ approaches to zero, ${\cal I}^{A E_1}$
approaches to ${\cal I}_{opt}$. This means that failure of the Eve1's optimal strategy is 
only due to the fact that Eve2 increases the disturbance. We plot ${\cal I}^{A E_2}$ in 
Fig. 4(b) when $D^{(1)} = 0.1$, $0.2$ and $0.3$ respectively. For comparison we plot
${\cal I}_{opt}$ and ${\cal I}^{AB}$ together. As expected ${\cal I}^{A E_2}$ approaches to 
${\cal I}_{opt}$ in the limit $D^{(1)} \rightarrow 0$. In this case, however, 
${\cal I}^{A E_2}$ decreases very rapidly compared to ${\cal I}^{A E_1}$ with increasing 
$D^{(1)}$. This seems to be mainly due to the fact that Eve2's information gain is affected by
Eve1 as shown in Eq.(\ref{2gain-1}).

\begin{figure}[ht!]
\begin{center}
\includegraphics[height=8.0cm]{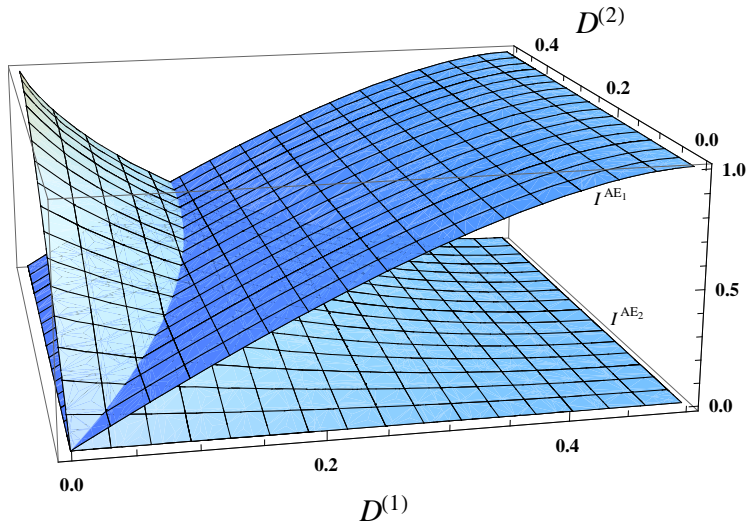}
\caption[fig5]{The $D^{(1)}$- and $D^{(2)}$-dependence of ${\cal I}^{A E_1}$ and 
${\cal I}^{A E_2}$.
In most region ${\cal I}^{A E_1}$ is larger than ${\cal I}^{A E_2}$. This seems to be mainly
due to the factor $1 - D_{xy}^{(1)}$ in Eq.(\ref{2gain-1}). However, in the small $D^{(1)}$ region
${\cal I}^{A E_2}$ becomes larger than ${\cal I}^{A E_1}$ because this multiplication factor 
becomes nearly unit in this region.}
\end{center}
\end{figure}

In Fig. 5 we plot ${\cal I}^{A E_1}$ and ${\cal I}^{A E_2}$ together as functions of 
$D^{(1)}$ and $D^{(2)}$. In most region ${\cal I}^{A E_1}$ is larger than ${\cal I}^{A E_2}$.
This is also due to the $D^{(1)}$-dependence of Eve2's information gain $G_{\lambda}^{(2)}$. 
In the small $D^{(1)}$ region, however, ${\cal I}^{A E_2}$ becomes larger than 
${\cal I}^{A E_1}$. This is due to the fact that Eve1 cannot gain information without increasing 
$D^{(1)}$ as Eq.(\ref{2gain-1}) indicates. 

\section{Three Eavesdroppers}

%%%%%%%%%%%%%%%%%%%%%%%%%%%%%%%%%%%%%%%%%%%%%%%%%%%%%%%%
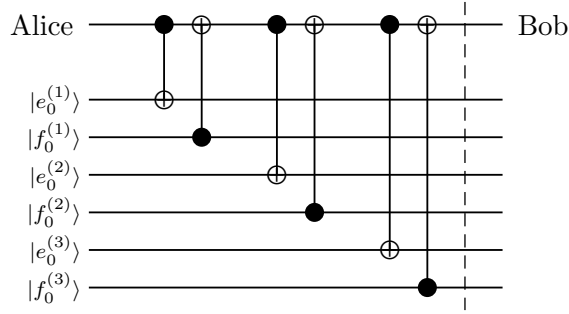
\begin{figure}
\begin{center}
\begin{pspicture}(0, 0)(7.5, 4.5)
%%%%%%%%%%  GATES

%%%%%%%%%%  Horizontal LINES
   \psline[linewidth=0.7pt]( 1.00,  4.00)( 6.50,  4.00)%
   
   \psline[linewidth=0.7pt]( 1.00,  3.00)( 6.50,  3.00)%
   \psline[linewidth=0.7pt]( 1.00,  2.50)( 6.50,  2.50)%
   
   \psline[linewidth=0.7pt]( 1.00,  2.00)( 6.50,  2.00)%
   \psline[linewidth=0.7pt]( 1.00,  1.50)( 6.50,  1.50)%
   
   \psline[linewidth=0.7pt]( 1.00,  1.00)( 6.50,  1.00)%
   \psline[linewidth=0.7pt]( 1.00,  0.50)( 6.50,  0.50)%
%%%%%%%%%%  Vertical LINES
   \psline[linewidth=0.7pt]( 2.00,  4.00)( 2.00,  3.00)%
   \psline[linewidth=0.7pt]( 2.50,  4.00)( 2.50,  2.50)%
   
   \psline[linewidth=0.7pt]( 3.50,  4.00)( 3.50,  2.00)%
   \psline[linewidth=0.7pt]( 4.00,  4.00)( 4.00,  1.50)%
   
   \psline[linewidth=0.7pt]( 5.00,  4.00)( 5.00,  1.00)%
   \psline[linewidth=0.7pt]( 5.50,  4.00)( 5.50,  0.50)%
   
   \psline[linewidth=0.5pt, linestyle=dashed]( 6.00,  4.30)( 6.00,  0.20)%
%%%%%%%%% DOTS
\psdot[dotstyle=*, dotscale=2]( 2.00,   4.00)%
   \psdot[dotstyle=*, dotscale=2]( 2.50,   2.50)%
   \psdot[dotstyle=oplus, dotscale=2]( 2.00,   3.00)%
   \psdot[dotstyle=oplus, dotscale=2]( 2.50,   4.00)%

   \psdot[dotstyle=*, dotscale=2]( 3.50,   4.00)%
   \psdot[dotstyle=*, dotscale=2]( 4.00,   1.50)%
   \psdot[dotstyle=oplus, dotscale=2]( 3.50,   2.00)%
   \psdot[dotstyle=oplus, dotscale=2]( 4.00,   4.00)%
   
   \psdot[dotstyle=*, dotscale=2]( 5.00,   4.00)%
   \psdot[dotstyle=*, dotscale=2]( 5.50,   0.50)%
   \psdot[dotstyle=oplus, dotscale=2]( 5.00,   1.00)%
   \psdot[dotstyle=oplus, dotscale=2]( 5.50,   4.00)%
%%%%%%%%%%  TEXTS
   \rput[r]( 0.80,  4.00){Alice}%
   \rput[l]( 6.70,  4.00){Bob}%
   
   \rput[r]( 0.90,  3.00){$\scriptstyle{\lvert e_{0}^{(1)} \rangle}$}
   \rput[r]( 0.90,  2.50){$\scriptstyle{\lvert f_{0}^{(1)} \rangle}$}
   
   \rput[r]( 0.90,  2.00){$\scriptstyle{\lvert e_{0}^{(2)} \rangle}$}
   \rput[r]( 0.90,  1.50){$\scriptstyle{\lvert f_{0}^{(2)} \rangle}$}

   \rput[r]( 0.90,  1.00){$\scriptstyle{\lvert e_{0}^{(3)} \rangle}$}
   \rput[r]( 0.90,  0.50){$\scriptstyle{\lvert f_{0}^{(3)}\rangle}$}
%%%%%%%%%%%%%%%%%%%%%%%%%%%%%%%%%%%%%

\end{pspicture}
\end{center}
\caption[fig6]{Quantum circuit expression for the situation where three eavesdroppers Eve1, Eve2, 
and Eve3 attack the usual BB84 protocol when Alice sends
a signal to Bob using $x-y$ basis. The top line belongs to Alice, 
next two lines to Eve1, next two lines to Eve2 and bottom two lines to Eve3. Time advances from
left to right.}
\end{figure}
%%%%%%%%%%%%%%%%%%%%%%%%%%%%%%%%%%%%%%%%%%%%%%%%%%%%%%%%%%%

In this section we consider a situation that three eavesdroppers called Eve1, Eve2, and Eve3
attack the usual BB84 protocol. As previous section we assume that they think they are 
unique eavesdroppers and choose their own symmetric strategies. Thus corresponding quantum 
circuit should be Fig. 6 when Alice sends a signal using $x-y$ basis. 

Using the compact notation used in Eq.(\ref{compact-1}), one can derive the entangled states
at the stage represented as a dotted line in Fig. 6. The final result becomes
\begin{eqnarray}
\label{compact-3}
& &\ket{x} \rightarrow \ket{X} = \sum_{i,j,k=0}^3 \alpha_i^{(1)} \alpha_j^{(2)} \alpha_k^{(3)} 
\ket{i+j+k}_2 \ket{i}_4 \ket{2i + j}_4 \ket{2i + 2j + k}_4            
                                                          \\     \nonumber
& &\ket{y} \rightarrow \ket{Y} = \sum_{i,j,k=0}^3 \alpha_i^{(1)} \alpha_j^{(2)} \alpha_k^{(3)} 
\ket{i+j+k+1}_2 \ket{i+2}_4 \ket{2i + j + 2}_4 \ket{2i + 2j + k + 2}_4.
\end{eqnarray}
Then, it is straightforward to construct the complete sets of the positive operators for 
eavesdroppers' POVM measurements. Following the similar calculational procedure, one can compute
the information gain for each eavesdropper. The final result can be summarized as following:
\begin{eqnarray}
\label{3gain-1}
& &G_{\lambda}^{(1)} = 1 - 2 \Delta^{(1)} = 2 \sqrt{D^{(1)} (1 - D^{(1)})}
                                                                    \\    \nonumber
& &G_{\lambda}^{(2)} = (1 - 2 \Delta^{(2)}) (1 - 2 D^{(1)}) = 
2 (1 - 2 D^{(1)}) \sqrt{D^{(2)} (1 - D^{(2)})}                    \\   \nonumber
& &G_{\lambda}^{(3)} = (1 - 2 \Delta^{(3)}) (1 - 2 D^{(1)}) (1 - 2 D^{(2)})= 
2 (1 - 2 D^{(1)}) (1 - 2 D^{(2)}) \sqrt{D^{(3)} (1 - D^{(3)})}.
\end{eqnarray}
Note that we remove all subscripts because they are not necessary in the symmetric strategy.
Eq.(\ref{3gain-1}) exhibits a simple pattern: the information gain for each eavesdropper is a 
multiplication of her own $1 - 2 \Delta$ factor with $1 - 2 D$ factor of other eavesdroppers
who perform their own strategies earlier. Using this rule, we can compute the information
gains when $n$ eavesdroppers attack with arbitrary number $n$ without explicit calculation. 
The mutual informations ${\cal I}^{A E_1}$, ${\cal I}^{A E_2}$ and ${\cal I}^{A E_3}$ 
reduce to
\begin{equation}
\label{3mutual-1}
{\cal I}^{A E_1} = \frac{1}{2} \phi \left( G_{\lambda}^{(1)} \right)  \hspace{1.0cm}
{\cal I}^{A E_2} = \frac{1}{2} \phi \left( G_{\lambda}^{(2)} \right)  \hspace{1.0cm}
{\cal I}^{A E_3} = \frac{1}{2} \phi \left( G_{\lambda}^{(3)} \right).
\end{equation}

Finally Bob's error rate $D_B$ can be read straightforwardly from Eq.(\ref{compact-3}):
\begin{equation}
\label{3Bob-1}
D_B = \left[ D^{(1)} \left(1 - D^{(2)}\right) + D^{(2)} \left(1 - D^{(1)}\right) \right]
      \left(1 - D^{(3)}\right) + 
      \left[ D^{(1)} D^{(2)} + \left(1 - D^{(1)}\right) \left(1 - D^{(2)}\right) \right]
      D^{(3)}.
\end{equation}
When $D^{(3)} = 0$, Eq.(\ref{3Bob-1}) exactly coincides with Eq.(\ref{2Bob-1}). If, furthermore,
$D^{(1)} = 0$ or $D^{(2)} = 0$, Eq.(\ref{3Bob-1}) reduces to Eq.(\ref{2Bob-1}) with changing
only Eve index.

\begin{figure}[ht!]
\begin{center}
\includegraphics[height=6.0cm]{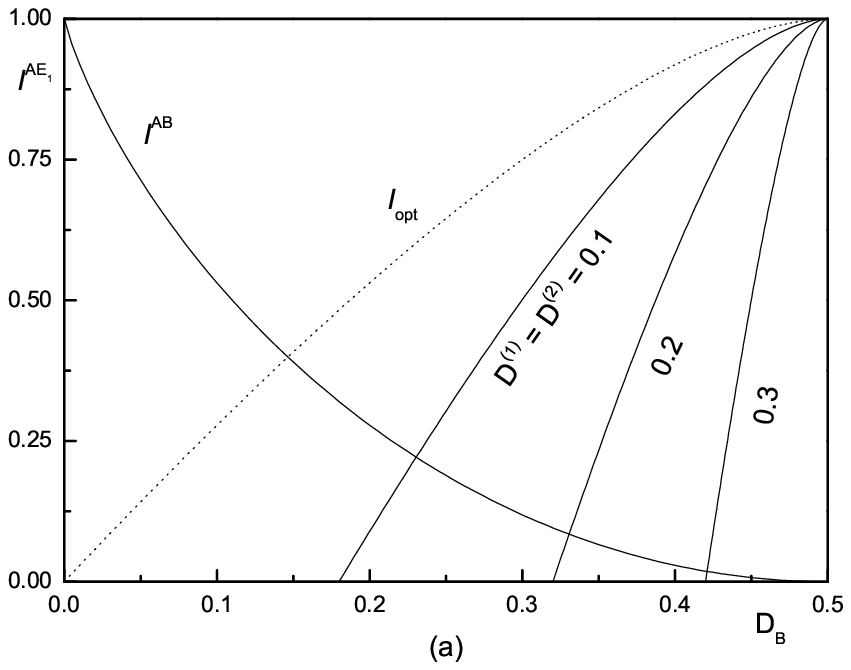}
\includegraphics[height=6.0cm]{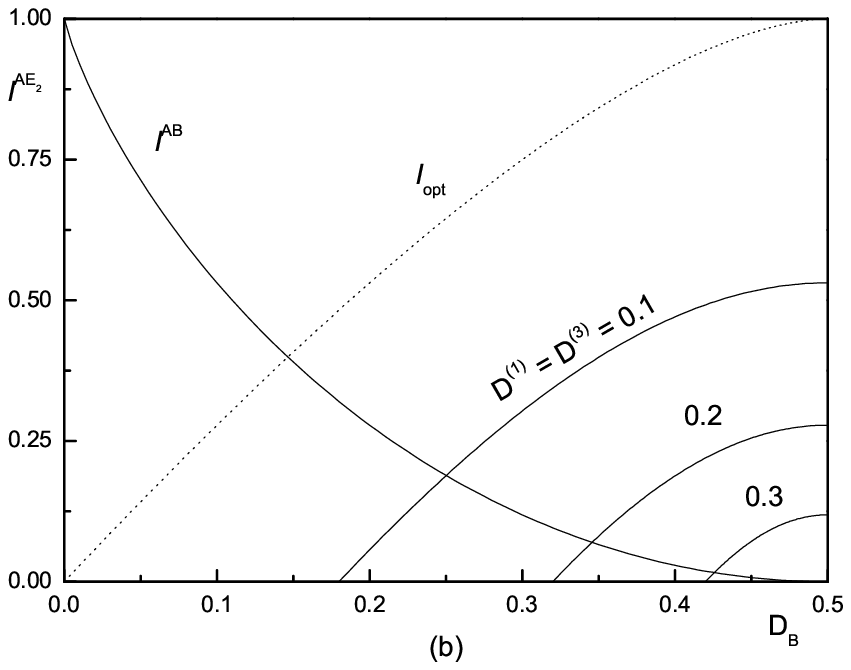}
\includegraphics[height=6.0cm]{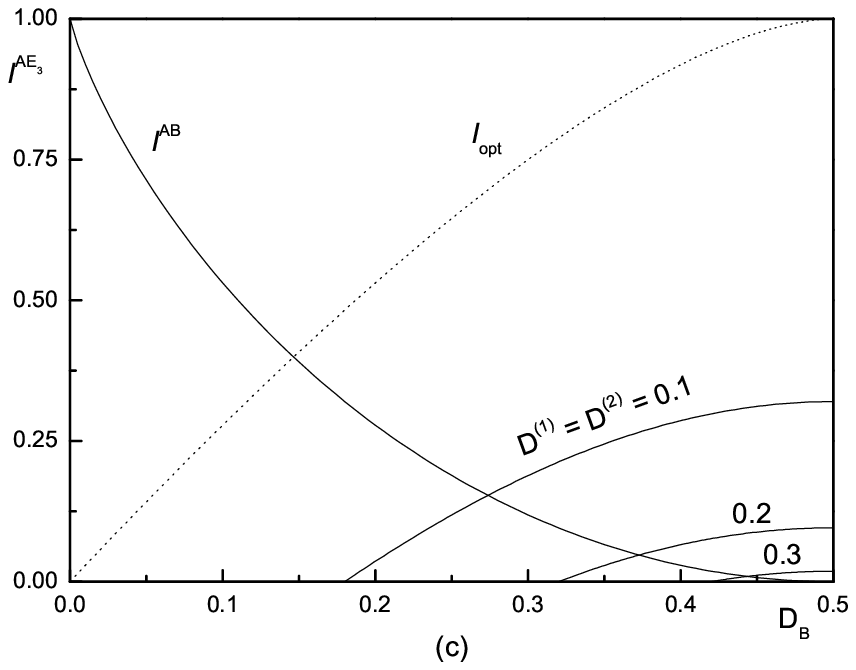}
\caption[fig7]{Plot of $D_B$-dependence of ${\cal I}^{A E_1}$ (Fig. 7(a)), ${\cal I}^{A E_2}$
(Fig. 7(b)), and ${\cal I}^{A E_3}$ (Fig. 7(c)). The optimal mutual information ${\cal I}_{opt}$
and Bob's mutual information ${\cal I}^{AB}$ are plotted together for comparison. This 
figure indicates that all optimal strategies performed by Eve1, Eve2, and Eve3 turn out fail.}
\end{center}
\end{figure}

Fig. 7 is the plot of $D_B$-dependence of ${\cal I}^{A E_1}$ (Fig. 7(a)), ${\cal I}^{A E_2}$ 
(Fig. 7(b)), and ${\cal I}^{A E_3}$ (Fig. 7(c)). We fixed $D^{(2)} = D^{(3)} = 0.1$, $0.2$
and $0.3$ in Fig. 7(a), $D^{(1)} = D^{(3)} = 0.1$, $0.2$ and $0.3$ in Fig. 7(b), and 
$D^{(1)} = D^{(2)} = 0.1$, $0.2$ and $0.3$ in Fig. 7(c). For comparison the optimal mutual 
information ${\cal I}_{opt}$ and Bob's information ${\cal I}^{AB}$ are plotted together. As 
Fig. 7 indicates, all optimal strategies turn out to fail. Especially, Eve3 gains very little
information compared to optimal one. This is mainly due to the fact that Eve1 and Eve2 disturb 
the Alice's signal before Eve3 starts her optimal strategy. Comparision of Fig. 7 with Fig. 4
indicates that mutual informations in the case of three eavesdroppers are overall less than
those in the case of two eavesdroppers. This seems to be due to the fact that Eve3's disturbance
of Alice's signal decreases ${\cal I}^{A E_1}$ and ${\cal I}^{A E_2}$ in the 
disturbance-information diagram.

\section{Conclusion}

%%%%%%%%%%%%%%%%%%%%%%%%%%%%%%%%%%%%%%%%%%%%%%%%%%%%%%%%
\begin{figure}
\begin{center}
\begin{pspicture}(0, 0)(8.0, 4.0)
%%%%%%%%%%  GATES

%%%%%%%%%%  Horizontal LINES
   \psline[linewidth=0.7pt]( 2.50,  3.50)( 7.00,  3.50)%
   
   \psline[linewidth=0.7pt]( 2.50,  2.50)( 7.00,  2.50)%
   \psline[linewidth=0.7pt]( 2.50,  2.00)( 7.00,  2.00)%
   
   \psline[linewidth=0.7pt]( 2.50,  1.00)( 7.00,  1.00)%
   \psline[linewidth=0.7pt]( 2.50,  0.50)( 7.00,  0.50)%
   
   \psline[linewidth=0.7pt, linestyle=dotted]( 4.50,  3.00)( 5.00,  3.00)%
%%%%%%%%%%  Vertical LINES
   \psline[linewidth=0.7pt]( 3.50,  3.50)( 3.50,  2.50)%
   \psline[linewidth=0.7pt]( 4.00,  3.50)( 4.00,  2.00)%
   
   \psline[linewidth=0.7pt]( 5.50,  3.50)( 5.50,  1.00)%
   \psline[linewidth=0.7pt]( 6.00,  3.50)( 6.00,  0.50)%
   
   \psline[linewidth=0.7pt, linestyle=dotted]( 3.00,  1.80)( 3.00,  1.20)%
   \psline[linewidth=0.7pt, linestyle=dotted]( 4.50,  1.80)( 4.50,  1.20)%
   \psline[linewidth=0.5pt, linestyle=dashed]( 6.50,  3.80)( 6.50,  0.20)%
%%%%%%%%% DOTS
   \psdot[dotstyle=*, dotscale=2]( 3.50,   3.50)%
   \psdot[dotstyle=*, dotscale=2]( 4.00,   2.00)%
   \psdot[dotstyle=oplus, dotscale=2]( 3.50,   2.50)%
   \psdot[dotstyle=oplus, dotscale=2]( 4.00,   3.50)%

   \psdot[dotstyle=*, dotscale=2]( 5.50,   3.50)%
   \psdot[dotstyle=*, dotscale=2]( 6.00,   0.50)%
   \psdot[dotstyle=oplus, dotscale=2]( 5.50,   1.00)%
   \psdot[dotstyle=oplus, dotscale=2]( 6.00,   3.50)%
%%%%%%%%%%  TEXTS
\rput[r]( 2.30,  3.50){Alice}%
   \rput[l]( 7.20,  3.50){Bob}%
   
   \rput[r]( 2.40,  2.25){$1^{st}$ eavesdropper $\bigg\{$}
   \rput[r]( 2.40,  0.75){$n^{th}$ eavesdropper $\bigg\{$}
%%%%%%%%%%%%%%%%%%%%%%%%%%%%%%%%%%%%%

\end{pspicture}
\end{center}
\caption[fig8]{Schematic diagram for the situation that $n$ eavesdroppers optimally attack the 
usual BB84 protocol performed by two trusted parties, Alice and Bob. We assume that Alice sends
a signal using $x-y$ basis. If Alice uses $u-v$ basis, this diagram should be modified by 
exchanging all control gates with target gates.}
\end{figure}
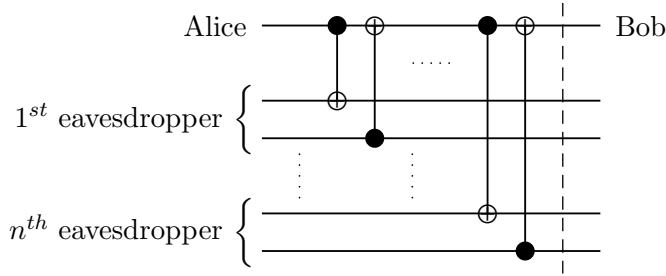
%%%%%%%%%%%%%%%%%%%%%%%%%%%%%%%%%%%%%%%%%%%%%%%%%%%%%%%%%%%

In this paper we have examined the situation that many eavesdroppers attack usual BB84 protocol
via their own symmetric optimal strategies. If the number of eavesdroppers is arbitrarily $n$
as shown in Fig. 8, Eq.(\ref{1optimal-7}), (\ref{2gain-1}) and (\ref{3gain-1}) 
imply that their information
gains are 
\begin{equation}
\label{ngain-1}
G^{(j)} = \left(1 - 2 \Delta^{(j)} \right) \left(1 - 2D^{(1)}\right) \cdots
                                   \left(1 - 2D^{(j-1)}\right) 
\hspace{1.0cm} (j=1, \cdots, n)
\end{equation}
and their mutual informations with Alice are
\begin{equation}
\label{nmutual-1}
{\cal I}^{(j)} = \frac{1}{2} \phi \left( G^{(j)} \right).
\hspace{1.0cm} (j=1, \cdots, n)
\end{equation}
Furthermore, Eq.(\ref{1optimal-11}), (\ref{2Bob-1}) and (\ref{3Bob-1}) imply that Bob's error
rate in the presence of $n$ eavesdroppers can be computed as follows. In order to distinguish
the number of eavesdroppers in the Bob's error rate, we use one more index such as 
$D_{B,j}$, which is Bob's error rate when $j$ eavesdroppers attack with symmetric optimal 
strategies. Then $D_{B,n}$ can be computed from $D_{B,n-1}$ by a recursion relation
\begin{equation}
\label{nBob-1}
D_{B,n} = D_{B,n-1} \left(1 - D^{(n)} \right) + 
D_{B, n-1}\bigg|_{D^{(n-1)} \rightarrow 1 - D^{(n-1)}} D^{(n)}.
\end{equation}
Since we know $D_{B,1}$ exactly, one can compute $D_{B,n}$ recursively.

Eq.(\ref{nmutual-1}) and (\ref{nBob-1}) enable us to plot the disturbance-information diagram
for any eavesdroppers. As commented already in the previous sections, all eavesdroppers' optimal
strategies cannot succeed eventually except very rare cases. Although the first eavesdropper can
obtain mutual information without disturbance arisen due to the other eavesdroppers, 
subsequent eavesdroppers increase
the Bob's error rate. This makes the mutual information of the first eavesdropper lower than the
optimal one in the disturbance-information diagram except $D^{(2)} = \cdots = D^{(n)} = 0$. 
The last
eavesdropper cannot gain information due to the disturbance of the Alice's signal arising due
to the previous eavesdroppings. Thus the last eavesdropper's optimal strategy fails except
$D^{(1)} = \cdots = D^{(n-1)} = 0$. Similar reasons make all optimal strategies fail.

It seems to be of interest to extend our results to the case of asymmetric eavesdropping. 
Probably it needs very long and tedious calculation. Furthermore, asymmetric eavesdropping 
strategy may be not important practically because Alice and Bob can notice the presence of 
eavesdropper more easily than the symmetric case. However, from the purely theoretical point
of view it is interesting issue because it may give origin of information gain and 
Bob's error rate. 

Although much attention has been paid to the optimal strategy in the various protocol, the 
properties of the non-optimal case are not examined sufficiently. Since, however, the effect
of decoherence makes it impossible for eavesdropper to perform the exactly optimal one, 
it seems to be more important to explore the strategies near to optimal from the aspect of the 
practical reason. Recently, it is found\cite{jung09-1} 
that the quantum resonance occurs in the Bob's 
error rate when Eve takes a near-optimal strategy. We believe that there are other new and 
interesting properties in the eavesdropping strategies near to optimal one. We would like to
explore this issue in the future.

{\bf Acknowledgement}: 
This work was supported by the Kyungnam University
Foundation Grant, 2008.

\end{document}